\pdfoutput=1
\documentclass[aps,pra,twocolumn,10pt]{revtex4-1}
\usepackage[utf8]{inputenc}
\usepackage{cochineal}
\usepackage[cochineal]{newtxmath}
\usepackage{mathtools,graphicx,microtype,siunitx}
\usepackage[cal=boondoxo]{mathalfa}
\usepackage[colorlinks=true,allcolors=blue]{hyperref}

\begin{document}

\title{The spin-orbit mechanism of electron pairing in quantum wires}

\author{Yasha Gindikin and Vladimir A.\ Sablikov}

\affiliation{Kotel'nikov Institute of Radio Engineering and Electronics,
Russian Academy of Sciences, Fryazino, Moscow District, 141190, Russia}

\begin{abstract}
We solve a two-body problem for electrons in a one-dimensional system to show that two-electron bound states can arise as a result of the image-potential-induced spin-orbit interaction (iSOI). The iSOI contributes an attractive component to the electron-electron interaction Hamiltonian that competes with the Coulomb repulsion and overcomes it under certain conditions. We find that there exist two distinct types of two-electron bound states, depending on the type of the motion that forms the iSOI: the relative motion or the motion of the electron pair as a whole. The binding energy lies in the meV range for realistic material parameters and is tunable by the gate potential.
\end{abstract}

\maketitle

\section{Introduction}

Electron pairing is commonly related to the attractive forces mediated by the crystal lattice~\cite{combescot2015excitons} or many-particle excitations~\cite{kagan2013modern}. In the present paper we propose a new electron pairing mechanism that stems from the electron motion and depends on their momentum.

Recently we have found that in the materials with the strong Rashba spin-orbit interaction a spin-dependent component appears in the pair electron-electron (e-e) interaction, which radically affects the electron system~\cite{PhysRevB.95.045138}.

In Refs.~\cite{PhysRevB.95.045138,doi:10.1002/pssr.201700256,2017arXiv170700316G} these effects have been studied for the spin-orbit interaction caused by the potential of image charges (iSOI) that electrons induce on a metallic gate placed nearby. The main result is the fact that the spin-dependent component of the e-e interaction Hamiltonian produced by the iSOI is attractive for  a particular spin orientation locked to momentum. This yields the dramatic consequences for the ground state and collective excitations of the many-electron system.

Thus, a one-dimensional (1D) electron system with sufficiently strong iSOI becomes unstable with respect to the electron-density-fluctuations, giving rise to the avalanche-like electrons inflow to the fluctuation region. When approaching the instability threshold, the charge stiffness of the electron system turns to zero, which reflects the mitigation of the Coulomb repulsion by the electron attraction owing to the iSOI~\cite{PhysRevB.95.045138}.

In this paper a two-body problem for electrons with iSOI is addressed. In contrast to the many-electron system, the two-electron problem allows for an exact solution and can answer the question of what effects the attracting interaction due to the iSOI leads to at any amplitude of the iSOI strength. We demonstrate that the iSOI component of the e-e interaction results in the electron pairing. We find that there exist two distinct types of two-electron bound states classified by the nature of the electron motion, owing to which the iSOI arises. 

The \textit{relative bound states} arise because of the reciprocal electron motion that creates an attractive potential for the relative motion of electrons with opposite spins. The magnitude of the attraction is set not only by the Coulomb forces between the electrons, but also by the electric field of the charged gate. This opens the possibility to tune the binding energy of the electron pair by changing the gate potential.

The \textit{convective bound states} appear as the center-of-mass motion creates an attractive potential for the pair of electrons with parallel spins. It is interesting that the attraction arises for electrons with a definite spin orientation that is locked to the direction of the center-of-mass momentum. The effective attraction grows with the center-of-mass momentum and the spin state of the pair depends on the momentum direction.

\begin{figure}[t]
	\includegraphics[width=0.9\linewidth]{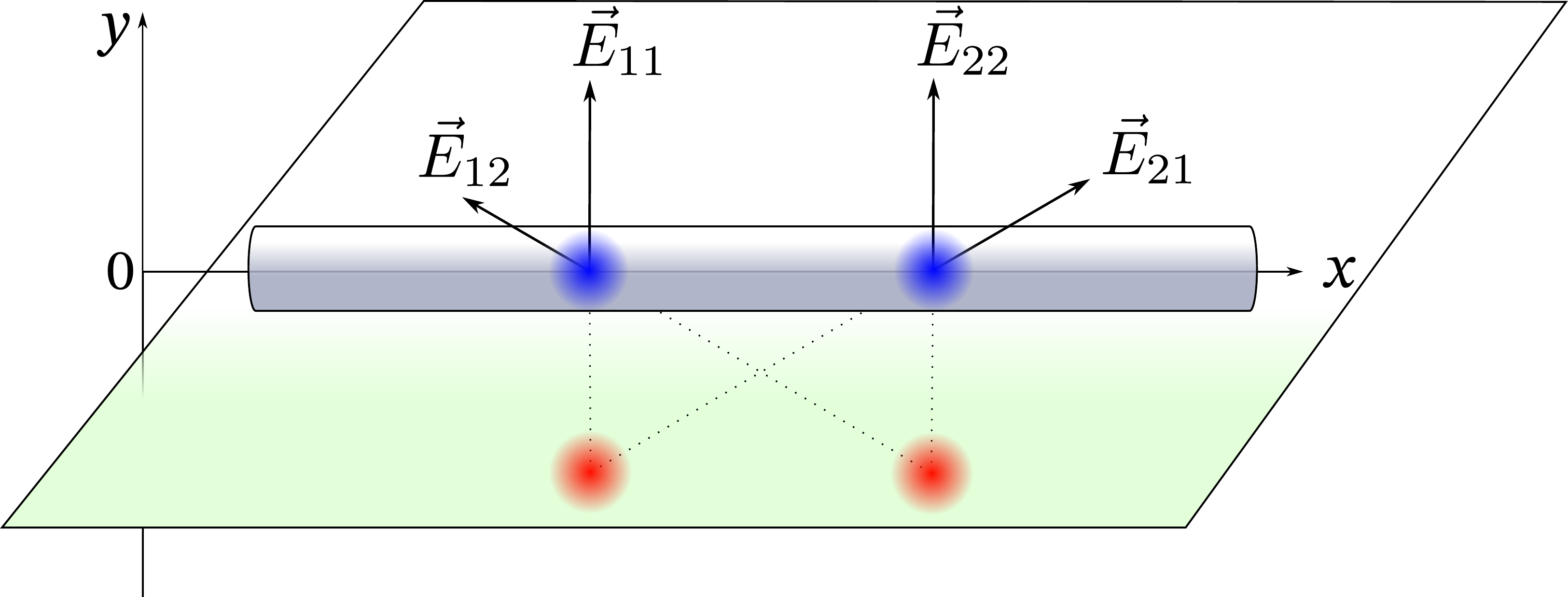}
		\caption{Two electrons in a quantum wire with image charges induced on a gate. The arrows show the electric fields acting on each electron from its own image as well as from the image of a neighboring electron.}
	\label{fig1}
\end{figure}

\section{The model}

Consider two electrons in a 1D quantum wire of a diameter $d$ parallel to the metallic gate situated in the $y = -a/2$ plane. The $x$ axis is directed along the wire as in Figure~\ref{fig1}.

A single-particle Hamiltonian is the sum of the kinetic energy and the Rashba SOI,
\begin{equation}
\label{1p}
	H = \sum_{i=1}^2 \frac{p_{x_i}^2}{2 m} + \frac{\alpha}{\hbar}\mathcal{F}p_{x_i}\sigma_{z_i} \, ,
\end{equation}
with $p_{x_i}$ being the $i$-th electron momentum, $\sigma_{z_i}$ the Pauli matrix, $\alpha$ the SOI constant, and $\mathcal{F} = e/ \epsilon a^2 + 2 \pi q_{\mathrm{g}}/\epsilon$ the $y$-component of the electric field that comes from the electron's own image and the background charge density $q_{\mathrm{g}}$ in the gate controlled by an external voltage.

The e-e interaction Hamiltonian has two parts. First, there is a Coulomb repulsion screened by the image charges. This one is described by the e-e interaction potential,
\begin{equation}
\label{Coul}
	\mathcal{U}(x_1-x_2) = \frac{e^2}{\epsilon \sqrt{{(x_1 - x_2)}^2 + d^2}} - \frac{e^2}{\epsilon \sqrt{{(x_1 - x_2)}^2 + a^2}}\,.
\end{equation}
The second part of the e-e interaction Hamiltonian is the SOI caused by the electric field $\vec{E}_{ij}$ acting on the $i$-th electron from the image of the other, $j$-th, electron,
\begin{equation}
\label{SOI}
	H_{\mathrm{iSOI}} = \frac{\alpha}{\hbar} \sum_{i \ne j} \frac{1}{2}[E_{ij}^{y} p_{x_i} + p_{x_i} E_{ij}^{y}]\sigma_{z_i}\,.
\end{equation}
The $y$-component of the field $\vec{E}_{ij}$ equals
\begin{equation}
	E_{ij}^{y} \equiv \mathcal{E}(x_i - x_j) = \frac{e a}{\epsilon{[{(x_i -x_j)}^2 + a^2]}^{\frac{3}{2}}}\,.
\end{equation}
We stress that the iSOI is essentially a two-particle interaction, in contrast to the commonly used one-particle Rashba Hamiltonian. The presence of the iSOI is a rather general property of low-dimensional structures since the image charges are induced not only in nearby conductors, but in a dielectric environment as well.

The two-electron wave function is a rank 4 spinor, $\Psi(x_1,x_2) = {(\psi_{\uparrow \uparrow},\psi_{\uparrow \downarrow},\psi_{\downarrow \uparrow},\psi_{\downarrow \downarrow})}^{\intercal}$. The full Hamiltonian~\eqref{1p}--\eqref{SOI} is diagonal in the corresponding basis, so the Schr\"odinger equation for $\Psi(x_1,x_2)$  splits into four separate equations for the spinor components. Prior to writing the equations let us switch from the coordinates of the individual electrons to the coordinate of relative motion $\xi = x_1 - x_2$ and the center-of-mass coordinate $\zeta = (x_1 + x_2)/2$.

The equations for $\psi_{\uparrow \downarrow}$ and $\psi_{\uparrow \uparrow}$ are
\begin{align}
\label{pot}
		\Bigl[ -\frac{\hbar^2}{m} \partial_{\xi}^2 -\frac{\hbar^2}{4m} \partial_{\zeta}^2 &- 2 i \alpha (\mathcal{F} + \mathcal{E}(\xi))\partial_{\xi}  \\
		&{} -  i\alpha \mathcal{E}'(\xi) + \mathcal{U}(\xi) \Bigr] \psi_{\uparrow \downarrow} = \varepsilon_{\uparrow \downarrow}  \psi_{\uparrow \downarrow} \notag
\end{align}
and
\begin{align}
\label{kin}
		&\Bigl[ - \frac{\hbar^2}{m} \partial_{\xi}^2 -\frac{\hbar^2}{4m} \partial_{\zeta}^2 - i \alpha (\mathcal{F} + \mathcal{E}(\xi))\partial_{\zeta} + \mathcal{U}(\xi)\Bigr]\psi_{\uparrow \uparrow}\notag\\
		&= \varepsilon_{\uparrow \uparrow} \psi_{\uparrow \uparrow}\,.
\end{align}
The equations for $\psi_{\downarrow \uparrow}$ and $\psi_{\downarrow \downarrow}$ are obtained from the above equations by changing the sign of $\alpha$. The solutions of the system are to be antisymmetrized with respect to the particle permutation.

\section{Relative bound states}

In Eq.~\eqref{pot} the reciprocal motion of electrons is separated from the center-of-mass motion. The wave function can be written as $\psi_{\uparrow \downarrow} = g(\zeta) f(\xi)$, where $g(\zeta)$ describes the free motion of the center-of-mass, $-\frac{\hbar^2}{4m} \partial_{\zeta}^2 g(\zeta) = (\varepsilon_{\uparrow \downarrow} - \varepsilon) g(\zeta)$, whereas the wave function of the reciprocal motion $f(\xi)$ satisfies the equation
\begin{align}
\label{potrel}
		&\left[- \frac{\hbar^2}{m} \partial_{\xi}^2 - 2 i \alpha (\mathcal{F} + \mathcal{E}(\xi))\partial_{\xi} - i\alpha \mathcal{E}'(\xi) + \mathcal{U}(\xi)\right] f(\xi) \notag\\
		&= \varepsilon  f(\xi)\,.
\end{align}
The gauge transformation $f(\xi) = u(\xi) e^{-i \phi(\xi)}$ with
\begin{equation}
\label{gauge}
	\phi(\xi)= \frac{m \alpha}{\hbar^2} \int_0^{\xi} (\mathcal{F} + \mathcal{E}(\eta))\,d \eta
\end{equation}
kills the first derivative to yield
\begin{equation}
\label{equ}
	- \frac{\hbar^2}{m} u'' + \left[\mathcal{U}(\xi) - \frac{m \alpha^2}{\hbar^2}{(\mathcal{F} + \mathcal{E}(\xi))}^2\right] u = \varepsilon u\,.
\end{equation}

\begin{figure}[t]
	\includegraphics[width=0.9\linewidth]{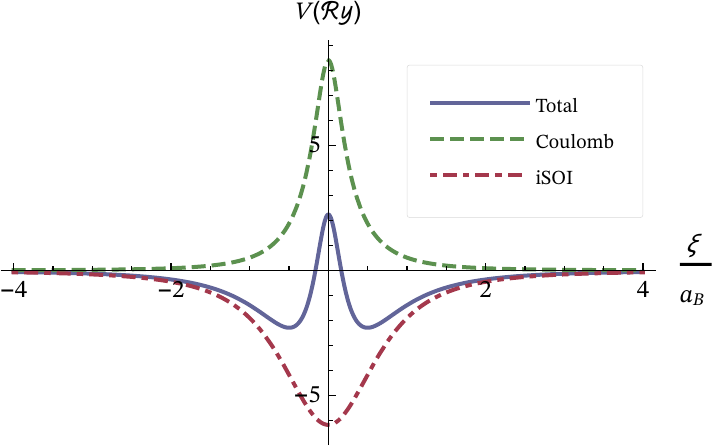}
		\caption{The effective potential profile $V(\xi)$ in the Eq.~\eqref{equ} of the relative motion and the contributions from the Coulomb e-e interaction and
		iSOI\@. The distance is normalized to Bohr's radius, the potential to the Rydberg constant in the material.}
	\label{fig2}
\end{figure}

Formally, this is a single-particle Schr\"odinger equation describing the motion in the potential profile of $V(\xi) = \mathcal{U}(\xi) - \frac{m \alpha^2}{\hbar^2}(\mathcal{E}^2(\xi) + 2 \mathcal{F} \mathcal{E}(\xi))$. The spatial profile of the potential is illustrated in Fig.~\ref{fig2}, with contributions from the Coulomb interaction and iSOI shown separately. This is clear that the Coulomb repulsion is suppressed by the iSOI\@. Moreover, the iSOI of a sufficient magnitude leads to the globally attractive potential $V(\xi)$, i.e.  $\int V(\xi)\, d\xi <0$. In 1D this suffices for a bound state to appear in the spectrum~\cite{simon1976bound}. The sufficient condition for the existence of a bound state is thus
\begin{equation}
\label{potcond}
	\tilde{\alpha}^2 > \frac{2 \log \frac{a}{d}}{\frac{3 \pi}{8}\frac{a_B^3}{a^3} + \frac{4 \epsilon a_B^3}{e a}\mathcal{F} }\,,
\end{equation}
with Bohr's radius $a_B=\epsilon \hbar^2/me^2$ and dimensionless SOI constant $\tilde{\alpha} = \alpha/ e a_B^2$. The fulfillment of this condition can be always achieved by increasing the field $\mathcal{F}$, that is by applying the potential to the gate.

In the case of zero gate potential one has $\mathcal{F} = e/\epsilon a^2$, so the condition~\eqref{potcond} becomes
\begin{equation}
\tilde{\alpha}^2 > \frac{2}{4 + \frac{3 \pi}{8}} {\left(\frac{a}{a_B}\right)}^3 \log \frac{a}{d}\,.
\end{equation}
A numerical estimate of this condition for the system based on a $\mathrm{Bi_2 Se_3}$, for which $\alpha \approx \SI{1300}{e \angstrom \squared}$~\cite{manchon2015new}, $a_B \approx \SI{52}{\angstrom}$ and hence $\tilde{\alpha} \approx 0.47$, gives the requirement of $a \le \SI{40}{\angstrom}$, which can be attained in modern nanostructures.

The binding energy is given by~\cite{landau1958course}
\begin{align}
	|\varepsilon| &= \frac{m}{4\hbar^2} {\left(\int_{-\infty}^{\infty} V(\xi)\,d \xi\right)}^2\\
	&= \frac{1}{2} Ry \cdot{\left[ \tilde{\alpha}^2 \left(\frac{3 \pi}{8}\frac{a_B^3}{a^3} + 4 \mathcal{F} \frac{\epsilon a_B^3}{ea}\right) - 2\log \frac{a}{d} \right]}^2\,,\notag
\end{align}
where $Ry = \hbar^2/2m a_B^2$ is the Rydberg constant in the material. Let us estimate the binding energy for the system based on $\mathrm{Bi_2 Se_3}$, this time assuming that the gate is biased. For reasonable values of the electric field $\mathcal{F} \approx \SI{3 e5}{V/cm}$, the distance to the gate $a \approx  \SI{50}{\angstrom}$ and the wire diameter $d \approx  \SI{10}{\angstrom}$, we get $|\varepsilon|$ of the order of $\SI{10}{meV}$.

Since the potential profile $V(\xi)$ is symmetric with respect to $\xi = 0$, the ground state is described by an even solution $u(\xi) = u(-\xi)$ of Eq.~\eqref{equ}. In other words, $u(\xi)$ is invariant under the permutation of electrons ($\xi \to -\xi$). Eq.~\eqref{gauge} shows that $\phi(\xi)$ is an odd function of $\xi$. Whence the antisymmetric two-electron wave function equals
\begin{equation}
	\Psi(x_1,x_2) = {\left( 0, e^{-i \phi(\xi)}, -e^{i \phi(\xi)}, 0 \right)}^{\intercal} u(\xi) g(\zeta)\,.
\end{equation}
The wave function of the relative bound state is seen to be of a mixed singlet-triplet type.

\section{Convective bound states}

The second type of the bound states, which we call convective, appears as the solution of Eq.~\eqref{kin}. Due to the translational invariance $\psi_{\uparrow \uparrow} = \exp (i K \zeta) f_K(\xi)$, with the wave function of the relative motion $f_K(\xi)$ defined by
\begin{align}
\label{kinrel}
		&\left[- \frac{\hbar^2}{m} \partial_{\xi}^2 + \left(\mathcal{U}(\xi) + \alpha K \mathcal{E}(\xi) \right) \right] f_K(\xi) \notag\\
		&= \left(\varepsilon_{\uparrow \uparrow} - \frac{\hbar^2 K^2}{4m} - \alpha K \mathcal{F}\right)  f_K(\xi)\,.
\end{align}

The most important feature of the convective bound states is that the binding potential $V(\xi) = \mathcal{U}(\xi) + \alpha K \mathcal{E}(\xi)$ depends on the center-of-mass momentum $K$, the sign and magnitude of which controls the existence or absence of the bound states as well as the binding energy. Large negative $K$ supports the existence of the convective bound states $\psi_{\uparrow \uparrow}$, while large positive $K$ supports $\psi_{\downarrow \downarrow}$. Thus the spin orientation of this purely triplet state is locked to the direction of $K$. In contrast to the relative bound states, the field $\mathcal{F}$ does not affect the potential profile and the binding energy, but only shifts the bottom of the conduction band.

Let us find the critical value of $K$ that allows for the appearance of a bound state. Note that the antisymmetric property of $\psi_{\uparrow \uparrow}$ requires that $f(\xi)$ be an odd function of $\xi$. Consequently, the Schr\"odinger equation for the zero-energy state can be solved on the half-axis,
\begin{equation}
\label{semi}
	\begin{dcases}
		-\partial_{\xi}^2 f + V(\xi) f = 0, \quad \xi \in (0, \infty)\\
		f \big|_{\xi = 0} = f \big|_{\xi = +\infty} = 0\,.
	\end{dcases}
\end{equation}
The transformation $r = \log \xi$, $u(r) = f(e^{r}) e^{-\frac{r}{2}}$ and $W(r) = e^{2r} V(e^r)$ maps Eq.~\eqref{semi} onto 
\begin{equation}
\label{trans}
	\begin{dcases}
		-\partial_{r}^2 u + W(r) u = -\frac{1}{4} u, \quad r \in (-\infty, \infty)\\
		u \big|_{r = \pm \infty} = 0\,.
	\end{dcases}
\end{equation}
Estimating the binding energy as $|\varepsilon| = \frac{1}{4} {\left[\int W(r)\,d r\right]}^2$, we arrive at the critical condition
\begin{equation}
	 \int_{-\infty}^{\infty} W(r)\, dr = -1\,.
\end{equation}
In terms of the original potential the criterion for the existence of the bound state takes the form
\begin{equation}
	\int_0^{\infty} \xi V(\xi)\, d \xi \leq -1\,,
\end{equation}
which is similar to the Bargmann limit on the number of bound states possessed by a central potential in three dimensions~\cite{Bargmann1952}. Finally, we obtain the desired condition for $K$,
\begin{equation}
	 \tilde{\alpha} K \ge (1 + a -d)a,
\end{equation}
with all variables normalized to Bohr's radius. Making an estimate for a system based on
$\mathrm{Bi_2 Se_3}$ with $a = \SI{30}{\angstrom}$ and $d = \SI{10}{\angstrom}$, we find that the convective bound state appears for $K \approx \SI{e7}{cm^{-1}}$.

\section{Conclusion and outlook}

We show that the image-potential-induced SOI gives an attractive contribution to the e-e interaction Hamiltonian that can overcome the Coulomb e-e repulsion under certain conditions. As a result, two electrons form bound states despite the Coulomb repulsion between them. The bound states can be of two types, depending on the nature of the motion due to which the spin-orbit interaction arises: the relative motion or the motion of the electron pair as a whole. In both cases the distance between the wire and the gate should be sufficiently small for the bound state to appear. The formation of the relative bound states is strongly facilitated by applying a gate voltage which allows one to tune their binding energy. In contrast, for the convective states it is important that the center-of-mass momentum is large enough, therefore they can be controlled by the current. The convective states have a purely triplet spin structure, whereas the relative states are formed by electrons with opposite spins. For realistic material parameters the binding energy can be in the meV range.

In the present paper we report the new mechanism of the electron pairing focusing on a two-body problem. Of course, there appears a more sophisticated and intriguing question of how the pairing manifests itself in a many-electron system. The problem is complicated since besides the electron pairing in a many-electron system there appears another strong effect due to the iSOI, namely, an instability of a homogeneous electron system with respect to the density fluctuations~\cite{PhysRevB.95.045138}. The relative role of both effects and their interplay are a challenging problem of further studies.

\textit{Acknowledgments.}---This work was partially supported by the Russian Foundation for Basic Research (Grant No 17--02--00309) and Russian Academy of Sciences.

\bibliography{paper}

\end{document}